\documentclass[a4paper,aps,prd,10pt,preprintnumbers,showpacs,twocolumn,superscriptaddress,nofootinbib,amsmath,amssymb]{revtex4-1}
\usepackage{graphicx}
\usepackage{cmap}
\usepackage[utf8]{inputenc}
\usepackage[T1]{fontenc}

\def\re#1{Re(#1)}
\def\im#1{Im(#1)}
\def\K{{\cal K}}
\def\Order#1{{\cal O}\left(#1\right)}

\begin{document}
\title{Grey-body factors for gravitational and electromagnetic perturbations around Gibbons-Maeda-Garfinkle-Horovits-Strominger black holes}
\author{Alexey Dubinsky}\email{dubinsky@ukr.net}
\affiliation{Pablo de Olavide University, Seville, Spain}
\begin{abstract}
While grey-body factors for a test scalar field in stringy black holes described by the renowned Gibbons-Maeda-Garfinkle-Horowitz-Strominger (GMGHS) solution have been analyzed in the literature, no such analysis exists for gravitons, likely due to the complexity of the perturbation equations. In this study, we utilize known data on quasinormal modes and the relationship between quasinormal modes and grey-body factors to derive these factors for gravitational and electromagnetic perturbations of the dilaton black hole. Our findings indicate that grey-body factors are significantly suppressed by the dilaton parameter as the black hole’s charge approaches its extreme value. The iso-spectrality between axial and polar channels of perturbations is broken in the presence of the dilaton field, which leads to different grey-body factors for different types of perturbations.  
\end{abstract}
\maketitle

\section{Introduction}

In black hole physics, the {\it grey-body factor} is a crucial concept for understanding radiation emitted through the Hawking radiation process \cite{Hawking:1975vcx}. It quantifies the modification of the radiation spectrum due to the potential barrier surrounding the black hole, which partially reflects some radiation back towards the black hole \cite{Page:1976df,Page:1976ki,Kanti:2004nr}.

Unlike idealized black-body radiation, where the spectrum is solely determined by temperature, the grey-body factor modifies the spectrum, accounting for the fact that not all radiation emitted near the event horizon reaches infinity. It represents the frequency-dependent scattering effects caused by the surrounding spacetime geometry.

Recently, a correspondence between quasinormal modes and grey-body factors has been identified for static asymptotically flat or de Sitter spherically symmetric black holes \cite{Konoplya:2024lir}, and it has also been verified for a broad class of rotating black hole spacetimes \cite{Konoplya:2024vuj}. While the correspondence is exact in the eikonal limit and shares similarities with the relationship between quasinormal modes and null geodesics \cite{Cardoso:2008bp,Konoplya:2017wot,Konoplya:2022gjp,Bolokhov:2023dxq}, including similar exceptions, it is only approximate beyond the eikonal regime. However, as shown in \cite{Konoplya:2024lir}, for Schwarzschild black holes and various generalizations, this correspondence is quite accurate not only for large but also for moderate values of the multipole numbers $\ell$. 

The particular spacetime of interest here is the renowned black hole solution in heterotic string theory, found independently by Gibbons and Maeda, and Garfinkle, Horowitz, and Strominger \cite{Gibbons:1987ps,Garfinkle:1990qj}. Heterotic string theory is a type of string theory that combines features of both bosonic and superstring theories, allowing for a unification of gravity with other fundamental forces. Heterotic strings are unique in that they incorporate different types of vibrations for the left-moving and right-moving modes of the string, enabling a rich structure that includes gauge fields. 

Perturbations and quasinormal modes for these black holes have been studied in several publications \cite{Holzhey:1991bx,Ferrari:2000ep,Malik:2024sxv,Konoplya:2001ji,Fernando:2003wc}. However, to the best of our knowledge, grey-body factors have only been analyzed for a test scalar field \cite{Koga:1995bs}. This limitation was likely due to the complexity of coupled gravitational, electromagnetic, and scalar perturbations, which were reduced to the Schrödinger wave-like form for both axial and polar channels in \cite{Ferrari:2000ep}.  

Here, we address this gap by calculating the grey-body factors for gravitational and electromagnetic perturbations, using the correspondence between grey-body factors and quasinormal modes. We demonstrate that the grey-body factors are strongly suppressed as the black hole’s charge increases.

The paper is organized as follows. In Sec.~II, we summarize the main information on the black hole metric and the wave equations for gravitational and electromagnetic perturbations. Sec.~III addresses the calculations of the grey-body factors through their correspondence with quasinormal modes. Finally, we conclude with a summary of the results obtained.

\section{The black hole metric and the wave equations}

The Gibbons-Maeda-Garfinkle-Horowitz-Strominger (GMGHS) black hole metric is an important solution in the study of dilaton gravity, describing a static, spherically symmetric, electrically charged black hole in theories that couple gravity, electromagnetic, and dilaton fields. This solution arises from an action that includes a dilaton field, and its properties depend on the dilaton coupling parameter \( a \). The metric generalizes the Reissner-Nordström solution by introducing a scalar field (dilaton) that modifies the spacetime structure, leading to novel characteristics relevant to string-inspired gravity theories and low-energy limits of heterotic string theory \cite{Gibbons:1987ps,Garfinkle:1990qj}. 

The line element for the GMGHS black hole with general dilaton coupling \( a \) is given by:
\begin{equation}
ds^2 = \lambda^2 dt^2 - \lambda^{-2} dr^2 - R^2 d\theta^2 - R^2 \sin^2 \theta \, d\phi^2,
\end{equation}
where
\begin{equation}
\lambda^2 = \left( 1 - \frac{r_+}{r} \right) \left( 1 - \frac{r_-}{r} \right)^{\frac{1-a^2}{1+a^2}},
\end{equation}
and
\begin{equation}
R^2 = r^2 \left( 1 - \frac{r_-}{r} \right)^{\frac{2a^2}{1+a^2}}.
\end{equation}

Here, \( r_+ \) and \( r_- \) denote parameters associated with the outer event horizon and the influence of the black hole charge, respectively. These parameters are related to the black hole mass \( M \) and electric charge \( Q \) by:
\begin{equation}
2M = r_+ + \frac{1 - a^2}{1 + a^2} \, r_-, \quad Q^2 = \frac{r_+ r_-}{1 + a^2}.
\end{equation}

The dilaton field \( \Phi \) and the electromagnetic field \( F_{\mu\nu} \) for this black hole solution are given by:
\begin{equation}
e^{2a\Phi} = \left( 1 - \frac{r_-}{r} \right)^{\frac{2a^2}{1+a^2}},
\end{equation}
\begin{equation}
F_{tr} = \frac{Q e^{2a\Phi}}{R^2}.
\end{equation}

The parameter \( a \) characterizes the coupling strength between the dilaton and electromagnetic fields, with different values representing distinct physical scenarios:
\begin{itemize}
    \item \( a = 0 \): The dilaton field decouples, reducing the metric to the classical Reissner-Nordström solution.
    \item \( a = 1 \): This value corresponds to the low-energy limit of heterotic string theory.
\end{itemize}

Here, we study the latter case of the black hole with string-theory-inspired corrections. Various properties and generalizations of these black holes have been investigated in \cite{Ghosh:1994wb,Koga:1995bs,Debnath:2015bna,Shiraishi:1992uuj,Shiraishi:1994txw,Shiraishi:1992ju,Maki:1992up,Cai:1993aa,Konoplya:2001ji,Jensen:1994ix,Chen:2009zzk,Ding:2020zgg}. Gravitational perturbations for dilatonic black holes have been analyzed in \cite{Holzhey:1991bx,Ferrari:2000ep}.

The perturbation equations are cumbersome and consist of chained equations. However, they can be diagonalized and reduced to the following two wave-like equations for \textit{axial} perturbations: 
\begin{equation}
\left( \frac{d^2}{dr_*^2} + \sigma^2 \right) \Psi_{a i} = V_{a i} \Psi_{a i} \quad (i = 1, 2),
\tag{36}
\end{equation}
and three wave-like equations for \textit{polar} perturbations:
\begin{equation}
\left( \frac{d^2}{dr_*^2} + \sigma^2 \right) \Psi_{p i} = V_{p i} \Psi_{p i} \quad (i = 1, 2, 3).
\tag{37}
\end{equation}

The explicit form of the effective potentials for polar perturbations is so complex that it was not even written down in \cite{Ferrari:2000ep}. When the scalar field is absent, the perturbations of the charged black hole can be divided into two independent channels: axial and polar, each containing two types of potentials, as two fields are perturbed simultaneously—gravitational and electromagnetic. When the coupled scalar field is included, a third type of wave equation is added to the polar channel, corresponding to the scalar field's degree of freedom.

Thus, the correspondence between the known quasinormal modes and grey-body factors provides an opportunity to determine the latter efficiently, despite the perturbation equations being extremely cumbersome and difficult to handle.

\vspace{4mm}
\section{Grey-body factors}

\begin{figure*}
\resizebox{\linewidth}{!}{\includegraphics{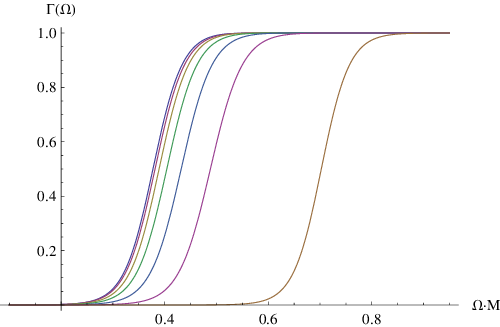}~~\includegraphics{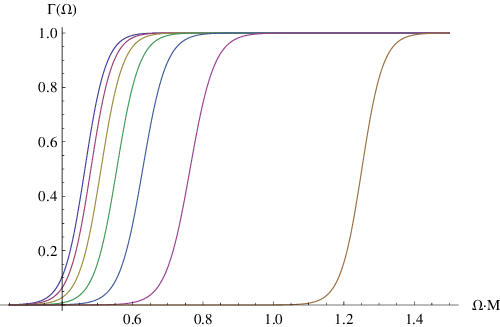}}
\caption{Grey-body factors for $\ell=2$ axial gravitational perturbations: $V_{1 a}$ (left), $V_{2 a}$ (right); the charge $Q =0.2$, $0.4$, $0.6$, $0.8$, $1$, $1.2$, $1.4$ from left to right.}\label{fig:GB1-2}
\end{figure*}

\begin{figure*}
\resizebox{\linewidth}{!}{\includegraphics{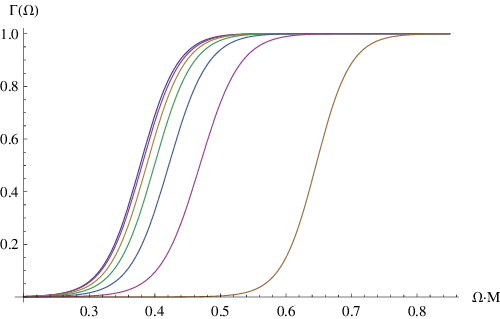}~~\includegraphics{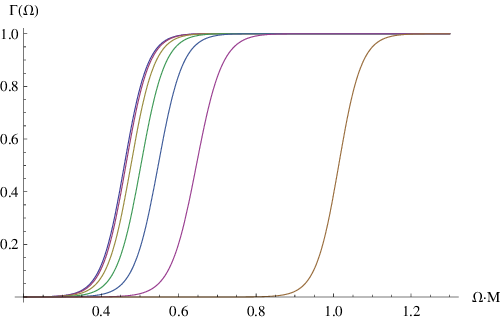}}
\caption{Grey-body factors for $\ell=2$ polar gravitational perturbations: $V_{1 p}$ (left) and $V_{2 p}$ (right); the charge $Q =0.2$, $0.4$, $0.6$, $0.8$, $1$, $1.2$, $1.4$ from left to right.}\label{fig:GB3-4}
\end{figure*}

\begin{figure*}
\resizebox{\linewidth}{!}{\includegraphics{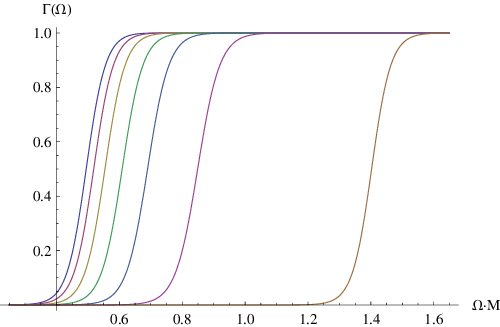}~~\includegraphics{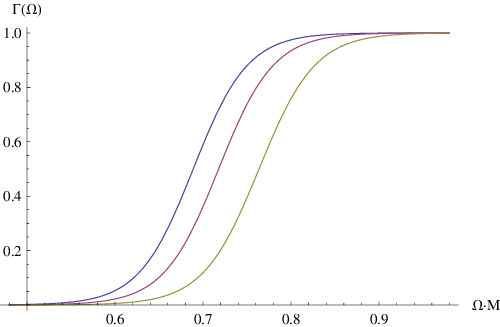}}
\caption{Grey-body factors for $\ell=2$ (left) and $\ell=3$ (right) polar gravitational perturbations: $V_{3 p}$; the charge $Q =0.2$, $0.4$, $0.6$ (for both cases), $0.8$, $1$, $1.2$, $1.4$ from left to right.}\label{fig:GB5and10}
\end{figure*}

%\begin{figure}
%\resizebox{\linewidth}{!}{\includegraphics{GBF10.eps}}
%\caption{Grey-body factors for $\ell=3$ polar gravitational perturbations: $V_{3 p}$; the charge $Q =0.2$, $0.4$, $0.6$ from %left to right.}\label{fig:GB10}
%\end{figure}

\begin{figure*}
\resizebox{\linewidth}{!}{\includegraphics{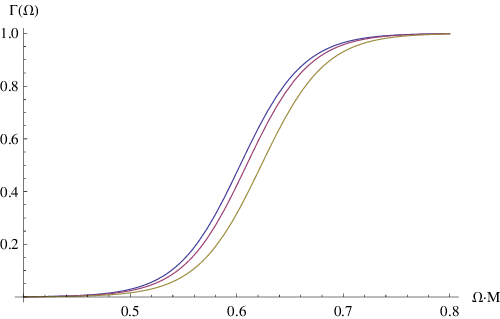}~~\includegraphics{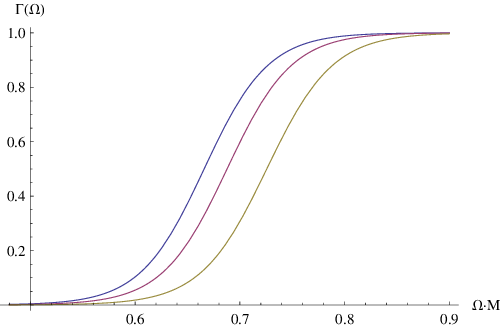}}
\caption{Grey-body factors for $\ell=3$ axial gravitational perturbations: $V_{1 a}$ (left), $V_{2 a}$ (right); the charge $Q =0.2$, $0.4$, $0.6$, from left to right.}\label{fig:GB6-7}
\end{figure*}

\begin{figure*}
\resizebox{\linewidth}{!}{\includegraphics{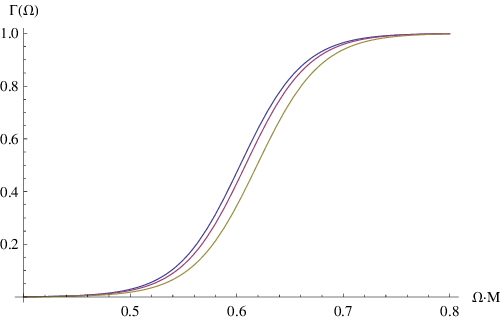}~~\includegraphics{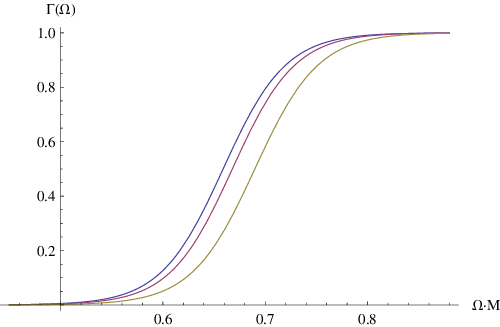}}
\caption{Grey-body factors for $\ell=3$ polar gravitational perturbations: $V_{1 p}$ (left) and $V_{2 p}$ (right); the charge $Q =0.2$, $0.4$, $0.6$ from left to right.}\label{fig:GB8-9}
\end{figure*}

The grey-body factor describes the probability that radiation emitted near the event horizon will reach a distant observer instead of being reflected back by the gravitational potential surrounding the black hole. This potential depends on the black hole's spacetime geometry.

The boundary conditions for solving the wave equation are based on the behavior of the field near the event horizon and at infinity:
\begin{enumerate}
    \item \textit{Near the Event Horizon} (\(r_* \rightarrow -\infty\)): The field is purely ingoing:
    \[
    \Psi(r_*) \sim T e^{-i \Omega r_*}.
    \]
    \item \textit{At Infinity} (\(r_* \rightarrow +\infty\)): The field is a combination of ingoing and outgoing waves:
    \[
    \Psi(r_*) \sim e^{-i \Omega r_*} + R e^{i \Omega r_*},
    \]
    where \( R \) is the reflection coefficient. These boundary conditions are used to determine the intensity of Hawking radiation \cite{Page:1976df,Page:1976ki,Kanti:2004nr,Harris:2003eg,Kokkotas:2010zd} or to analyze various radiation phenomena near black holes, such as superradiance \cite{1971JETPL..14..180Z,Starobinskil:1974nkd,Starobinsky:1973aij,Bekenstein:1998nt,East:2017ovw,Hod:2012zza,Konoplya:2008hj,Konoplya:2014lha}.
\end{enumerate}

The \textit{grey-body factor} $\Gamma_{\ell}(\omega)$ is then associated with the transmission coefficient:
$$
\Gamma_{\ell}(\Omega) = \left|T_\ell (\Omega)\right|^2.
$$

Grey-body factors, as well as quasinormal modes, can be calculated using the higher-order WKB (Wentzel-Kramers-Brillouin) approach \cite{Schutz:1985km,Iyer:1986np,Konoplya:2003ii,Matyjasek:2017psv}, which is based on the Taylor expansion of the wave function near the peak of the effective potential and matching it with the WKB series in asymptotic regions. This approach is widely used \cite{Dubinsky:2024nzo,Dubinsky:2024rvf,Dubinsky:2024aeu,Dubinsky:2024hmn,Malik:2024tuf,Abdalla:2005hu,Bolokhov:2024ixe,Bolokhov:2023dxq,Konoplya:2005sy,Skvortsova:2024wly,Skvortsova:2024atk,Konoplya:2006ar,Cuyubamba:2016cug,Xiong:2021cth,Momennia:2022tug,Kodama:2009bf}, and is known to be sufficiently accurate for $\ell > s$ perturbations. The formulas were derived for spherically symmetric and asymptotically flat black holes via the well-known WKB expression for grey-body factors,
\begin{equation}\label{eq:gbfactor}
\Gamma_{\ell}(\Omega) = \dfrac{1}{1+e^{2\pi i \K}},
\end{equation}
where $\K$ depends on the values of the effective potential and its derivatives up to the $2i$-th order for the $t$-th order of the WKB method \cite{Schutz:1985km,Iyer:1986np,Konoplya:2003ii,Matyjasek:2017psv}. The above WKB formula has been applied in the calculation of grey-body factors for black holes and wormholes in various studies \cite{Konoplya:2010kv,Konoplya:2020jgt,Stashko:2024wuq,Bolokhov:2024voa,Dubinsky:2024nzo,Fernando:2016ksb,Konoplya:2023ahd}.

According to \cite{Konoplya:2024lir}, the function $\K$ can be found in an alternative way, due to its correspondence with the fundamental mode and first overtone of the quasinormal spectrum. Thus, we have:
\begin{widetext}
\begin{eqnarray}\nonumber
&& i\K = \frac{\Omega^2 - \re{\omega_0}^2}{4\re{\omega_0}\im{\omega_0}}\Biggl(1 + \frac{(\re{\omega_0} - \re{\omega_1})^2}{32\im{\omega_0}^2}
%\\\nonumber&&\qquad\qquad
-\frac{3\im{\omega_0} - \im{\omega_1}}{24\im{\omega_0}}\Biggr)
-\frac{\re{\omega_0} - \re{\omega_1}}{16\im{\omega_0}}
\\\nonumber&& -\frac{(\omega^2 - \re{\omega_0}^2)^2}{16\re{\omega_0}^3\im{\omega_0}}\left(1 + \frac{\re{\omega_0}(\re{\omega_0} - \re{\omega_1})}{4\im{\omega_0}^2}\right)
%\\\nonumber&&
+\frac{(\omega^2 - \re{\omega_0}^2)^3}{32\re{\omega_0}^5\im{\omega_0}}\Biggl(1 + \frac{\re{\omega_0}(\re{\omega_0} - \re{\omega_1})}{4\im{\omega_0}^2}
\\\nonumber&&\qquad + \re{\omega_0}^2\Biggl(\frac{(\re{\omega_0} - \re{\omega_1})^2}{16\im{\omega_0}^4}
%\\&&\qquad\qquad 
-\frac{3\im{\omega_0} - \im{\omega_1}}{12\im{\omega_0}}\Biggr)\Biggr) + \Order{\frac{1}{\ell^3}}.
\label{eq:gbsecondorder}
\end{eqnarray}
\end{widetext}
Here $\omega_0$ and $\omega_1$ are the fundamental mode and the first overtone, respectively.

Grey-body factors modify the idealized Hawking radiation spectrum. For small black holes near the Planck scale, transmission through the potential barrier becomes negligible at low frequencies, and grey-body factors significantly reduce the radiation reaching infinity.

It turns out that larger charge \( Q \) leads to smaller grey-body factors. This can be explained by the behavior of quasinormal modes: as numerical data in \cite{Ferrari:2000ep} show, a larger charge \( Q \) increases the real part of the quasinormal frequency while the imaginary part remains almost unchanged. Therefore, according to the correspondence in its eikonal limit
\begin{equation}
i\K = \frac{\Omega^2 - \re{\omega_0}^2}{4\re{\omega_0}\im{\omega_0}},
\end{equation}
a larger $Re(\omega_{0})$  yields smaller grey-body factors. Another explanation is that for larger charges, the effective potential height increases, leading to a greater fraction of particles being reflected back and a lower transmission coefficient. This dependence of quasinormal modes on the charge is unusual in the absence of the dilaton, where the real oscillation frequency might decrease with charge, leading to larger grey-body factors \cite{Konoplya:2007zx}.

Another important difference between the grey-body factors of Reissner–Nordström black holes and those of the dilatonic black hole is the breakdown of iso-spectrality in the latter case. As shown in Figures \ref{fig:GB1-2}-\ref{fig:GB8-9}, the grey-body factors differ significantly between axial and polar perturbations.

In addition to the dilaton, string theory corrections include higher-curvature terms, the most dominant of which is the Gauss-Bonnet term. Consequently, black hole spacetime can be derived within Einstein-dilaton-Gauss-Bonnet theory \cite{Kanti:1995vq}. Gravitational perturbations in this case are even more complex, especially given that the metric itself is determined numerically \cite{Kanti:1995vq}. However, grey-body factors cannot be obtained for this case through correspondence with quasinormal modes, as the analogous correspondence with null geodesics breaks down \cite{Konoplya:2019hml}.

\section{Conclusions}

Perturbations of the gravitational field of a charged dilaton black hole involve complex perturbations of coupled electromagnetic and scalar fields \cite{Holzhey:1991bx,Ferrari:2000ep}, making the calculation of grey-body factors a non-trivial problem. However, once this technically challenging work was accomplished for quasinormal modes \cite{Ferrari:2000ep,Konoplya:2001ji}, one can employ the correspondence between quasinormal modes and grey-body factors developed in \cite{Konoplya:2024lir,Konoplya:2024vuj}. We utilized this correspondence to address a gap in the literature on this classical solution and computed the grey-body factors for black holes with string theory corrections. It turns out that grey-body factors are significantly suppressed when the charge (coupled to the dilaton field) is introduced.

We also observe that the axial and polar perturbations are no longer iso-spectral and differ considerably from each other.

Another recent and interesting aspect of grey-body factors in black holes is their greater stability under small static spacetime deformations \cite{Rosato:2024arw,Oshita:2024fzf} compared to the overtones of the quasinormal spectrum, which are highly sensitive to near-horizon deformations \cite{Konoplya:2022pbc,Konoplya:2022hll}. Additionally, the profile of a gravitational wave in the high-frequency regime can be fitted using grey-body factors \cite{Oshita:2023cjz,Okabayashi:2024qbz}, which appears to reflect the correspondence found in \cite{Konoplya:2024lir,Konoplya:2024vuj}. Consequently, the grey-body factors derived here could be used not only for accurately calculating the intensity of Hawking radiation but may also be applicable to the description of gravitational waves around dilaton black holes.

\vspace{4mm}
\begin{acknowledgments}
The author would like to thank R. A. Konoplya for useful discussions and careful reading of the manuscript. The author acknowledges the Pablo de Olavide University for their support through the University-Refuge Action Plan. 
\end{acknowledgments}

\bibliography{bibliography}

\end{document}